\documentclass[fleqn,10pt]{wlscirep}

\newcommand{\new}[1]{\textcolor{black}{#1}}

\newcommand{\ket}[1]{\lvert {#1} \rangle \,}

\newcommand{\bra}[1]{\langle {#1} \rvert \,}

\newcommand{\braket}[2]{\langle {#1} | {#2} \rangle \,}
\newcommand{\mean}[3]{\langle {#1} | {#2} | {#3} \rangle \,}
\newcommand{\mv}[1]{\langle {#1} \rangle \,}

\newcommand{\Tr}[1]{\mathrm{Tr}\!\left[#1\right]}

\newcommand{\af}{\hat{a}_{_{\!f}}}
\newcommand{\Df}{\mathcal{D}_{_{\!f}}}
\newcommand{\gf}{\gamma_{_{\!f}}}

\title{Exact results for Schr\"odinger cats\\in driven-dissipative systems and\\ their feedback control}

\author[1]{Fabrizio Minganti}
\author[1]{Nicola Bartolo}
\author[1]{Jared Lolli}
\author[1]{Wim Casteels}
\author[1,*]{Cristiano Ciuti}
\affil{Universit\'{e} Paris Diderot, Sorbonne Paris Cit\'{e}, Laboratoire Mat\'{e}riaux et Ph\'{e}nom\`{e}nes Quantiques,
	CNRS-UMR7162, 75013 Paris, France}
\affil[*]{cristiano.ciuti@univ-paris-diderot.fr}

\begin{abstract}
\new{In quantum optics, photonic Schr\"odinger} cats are superpositions of two coherent states with opposite phases and with a significant number of photons. 
Recently, these states have been observed in the transient dynamics of driven-dissipative resonators subject to engineered two-photon processes.
Here we present an exact analytical solution of the steady-state density matrix for this class of systems, including one-photon losses, which are considered detrimental for the achievement of cat states. 
We demonstrate that the unique steady state is a statistical mixture of two cat-like states with opposite parity, in spite of significant one-photon losses. 
The transient dynamics to the steady state depends dramatically on the initial state and can pass through a metastable regime lasting orders of magnitudes longer than the photon lifetime.
By considering individual quantum trajectories in photon-counting configuration, we find that the system intermittently jumps between two cats. \new{Finally, we propose and study a feedback protocol based on this behaviour to generate a pure cat-like steady state.}
\end{abstract}
\begin{document}

\flushbottom
\maketitle
\thispagestyle{empty}

\section*{Introduction}

Quantum nonlinear optical systems are an invaluable tool to explore the quantum world and its striking features~\cite{HarocheBOOK}.
Generally, these systems are out-of-equilibrium:  photons must be continuously pumped into the system to replace those inevitably dissipated.
Effective photon-photon interactions can be mediated by an active medium, such as atoms or excitons in cavity QED or Josephson junctions in circuit QED resonators~\cite{CarusottoRMP13}.
The concepts of reservoir engineering~\cite{PoyatosPRL96,VerstraeteNatPhys09,TanPRA13,LinNature13,AsjadJPB14,RoyPRA15} and feedback~\cite{BrunePRA92, SlosserPRL95,SayrinNature11,ZhouPRL12,ShankarNature13,HeinPRA15} emerged to stabilize nonclassical steady states \new{in photonic and optomechanical resonators}. In particular, new opportunities arise via engineering of two-photon pumping and/or two-photon dissipation ~\cite{ArenzJPB13,MirrahimiNJP14}.

\new{Since their theoretical conception~\cite{SchrodingerN23}, Schr\"odinger's cats have captured the collective imagination, because they are non-classical states at the macroscopic level.}
In quantum optics, the states of the electromagnetic field closest to the classical ones are the coherent states \mbox{$\ket{\alpha}=e^{-|\alpha|^2/2}\,\sum_n (n!)^{-1/2}\,\alpha^n\,\ket{n}$}, having a well-defined mean amplitude $\vert \alpha \vert$ and phase \new{(being $\ket{n}$ the $n$-photon Fock state)}.
Photonic Schr\"odinger cat states are a quantum superposition of coherent states $\ket{\alpha}$ and $\ket{-\alpha}$~\cite{HarocheBOOK,GerryJOMO93, GillesPRA94}: 
\begin{equation}\label{eq:catdef}
	\ket{\mathcal{C}_\alpha^\pm}=
	\frac{\ket{\alpha}\pm \ket{-\alpha}}
	{\sqrt{2\,(1\pm e^{-2|\alpha|^2})}}.
\end{equation}
Contrarily to the coherent states, $\ket{\mathcal{C}_\alpha^\pm}$ are eigenstates of the photon-parity operator  $\hat{\mathcal{P}}=e^{i\pi\hat{n}}$, with $\hat{n}=\hat{a}^\dagger\hat{a}$ the photon number operator.
In fact, they are a superposition of only even (odd) number states.
Two-photon processes are known to drive the system towards this kind of \new{photonic} states~\cite{GerryJOMO93,GillesPRA94}.
However, one-photon losses are unavoidable even in the best resonators.
As a result, the presence of both one- and two-photon dissipations
makes the life of the cat states more intriguing~\cite{GillesPRA94,ZurekRMP03,ArenzJPB13,LeghtasScience15}. 

In this work, we provide an exact analytical solution for the steady state of this class of systems.
We show that the rich transient dynamics depends dramatically on the initial state.
It can exhibit metastable plateaux lasting several orders of magnitude longer
than the single-photon lifetime.
We demonstrate that\new{, for a wide range of parameters around typical experimental ones~\cite{LeghtasScience15},} the unique steady-state density matrix has as eigenstates two cat-like states even for significant one-photon losses, with all the other eigenstates having negligible probability.
The study of individual trajectories reveals that, under photon counting on the environment, the system jumps between the two cat states, a property suggesting a feedback scheme to create pure cat-like steady states.

\section*{Results}

\begin{figure}[t]
	\includegraphics[width=0.6\textwidth]{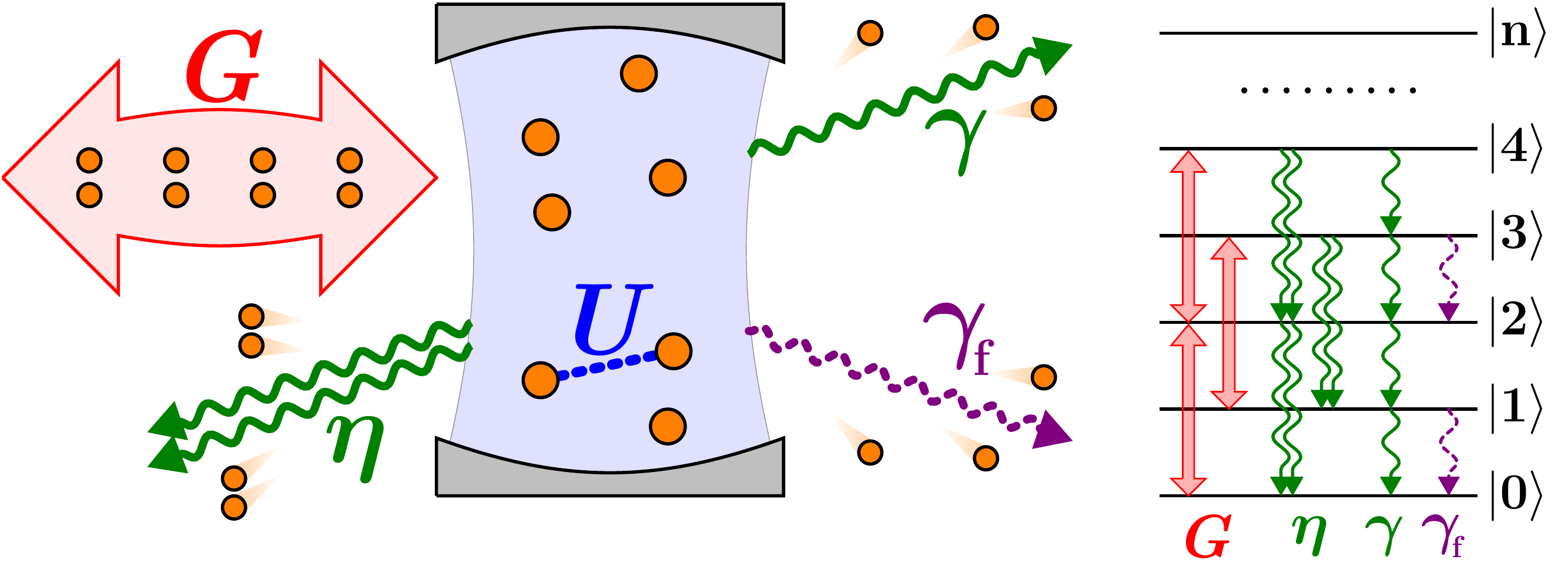}
	\caption{{\bf A sketch of the considered class of systems.}
		The picture represents a photon resonator subject to one-photon losses with rate $\gamma$.
		Via an engineered reservoir, the system is coherently driven by a two-photon pump with amplitude $G$ and has two-photon losses with rate $\eta$.
		We also consider an additional selective dissipation channel with rate $\gf$ acting only on odd states.
		$U$ quantifies the strength of the photon-photon interaction.
		On the right, we sketch the effects of the previous mechanisms on the Fock (number) states $\vert n \rangle$.}
	\label{fig:syst}
\end{figure}

\subsection*{Theoretical framework and analytic solution}
To start our treatment, we consider the master equation for the density matrix.
For a  system interacting with a Markovian reservoir, the time evolution of the reduced density matrix $\hat{\rho}$ is captured by the Lindblad master equation \mbox{$\partial_t \hat{\rho}= i/\hbar\,[\hat{\rho}, \hat{H}]+\mathcal{D} \hat{\rho}$}. The operator $\hat{H}$ is the Hamiltonian, while $\mathcal{D}$ is the Lindblad dissipation super-operator~\cite{CarmichaelBOOK,HarocheBOOK}.
In the frame rotating at the pump frequency, 
\begin{align}
	\label{eq:hamiltonian2phot_a}
	&\hat{H}=-\Delta\hat{a}^\dagger \hat{a}+\frac{U}{2}\hat{a}^\dagger \hat{a}^\dagger \hat{a} \hat{a} + \frac{G}{2} \hat{a}^\dagger \hat{a}^\dagger +  \frac{G^*}{2} \hat{a}  \hat{a}, \\
	\label{eq:hamiltonian2phot_b}             
	& \mathcal{D}_1\, \hat{\rho}=\frac{\gamma}{2}\left(2 \hat{a} \hat{\rho} \hat{a}^{\dagger} -\hat{a}^{\dagger}\hat{a} \hat{\rho} - \hat{\rho} \hat{a}^{\dagger} \hat{a} \right),  \\
	\label{eq:hamiltonian2phot_c}
	&\mathcal{D}_2\, \hat{\rho}=  \frac{\eta}{2}\left(2 \hat{a} \hat{a} \hat{\rho} \hat{a}^{\dagger}  \hat{a}^{\dagger} -\hat{a}^{\dagger}\hat{a}^{\dagger}\hat{a} \hat{a} \hat{\rho} - \hat{\rho} \hat{a}^{\dagger} \hat{a}^{\dagger}\hat{a}  \hat{a} \right),
\end{align}
where $\Delta$ is the pump-cavity detuning and $U$ the photon-photon interaction strength. $G$ is the amplitude of the two-photon pump, while $\gamma$ and $\eta$ represent, respectively, the one- and two-photon dissipation rates (see Fig.~\ref{fig:syst}).
The Lindblad dissipator $\mathcal{D} = \mathcal{D}_1 + \mathcal{D}_2$ is the sum of one- and two-photon loss contributions.
This model has been investigated theoretically~\cite{WolinskyPRL88, GillesPRA94,KrippnerPRA94,MirrahimiNJP14,EverittFICT14} and implemented experimentally~\cite{LeghtasScience15}.

In order to find a general and analytic solution for the steady-state density matrix $\hat{\rho}_{\rm ss}$, we used the formalism of the complex $P$-representation~\cite{DrummondJPA80}. Details about the derivation of our solution are in the Methods section. In spite of the several parameters in the model, our solution depends only on two dimensionless quantities, namely  $c=(\Delta+i\,\hbar\,\gamma/2)/(U-i\,\hbar\,\eta)$ and $g=G/(U-i\,\hbar\,\eta)$.
The former can be seen as a complex single-particle detuning $\Delta+i\,\hbar\,\gamma/2$ divided by a complex interaction energy $U-i\,\hbar\,\eta$; $g$ is instead the two-photon pump intensity normalized by the same quantity.
Hence, our exact solution for the steady-state density matrix elements in the Fock basis reads
\begin{equation} \label{eq:Rhonm}
		\bra{n}\hat{\rho}_{\rm ss}\ket{m}=
		\frac{1}{\mathcal{N}}\,
		\sum_{\ell=0}^{\infty} \frac{1}{\ell!\sqrt{n!\,m!}}\,
		\mathcal{F}(g,c,\ell+n)\,\mathcal{F}^*(g,c,\ell+m),
\end{equation}
where $\mathcal{N}$ is the normalization factor, chosen such that $\Tr{\hat{\rho}_{\rm ss}}=1$.
$\mathcal{F}(g,c,\ell)=(i\,\sqrt{g})^{\ell} \; _2F_1(-\ell,-c; -2c; 2)$, $_2F_1$ being the Gaussian hypergeometric function. 
Notably, $\mathcal{F}(g,c,\ell)=0$ for $\ell$ odd, meaning that, for \emph{any} finite value of the system parameters, there will be no even-odd coherences in the steady state.
\new{In what follows following, all the quantities marked with a tilde will refer to steady-state values.}

\begin{figure}[t]
	\includegraphics[width=0.93\textwidth]{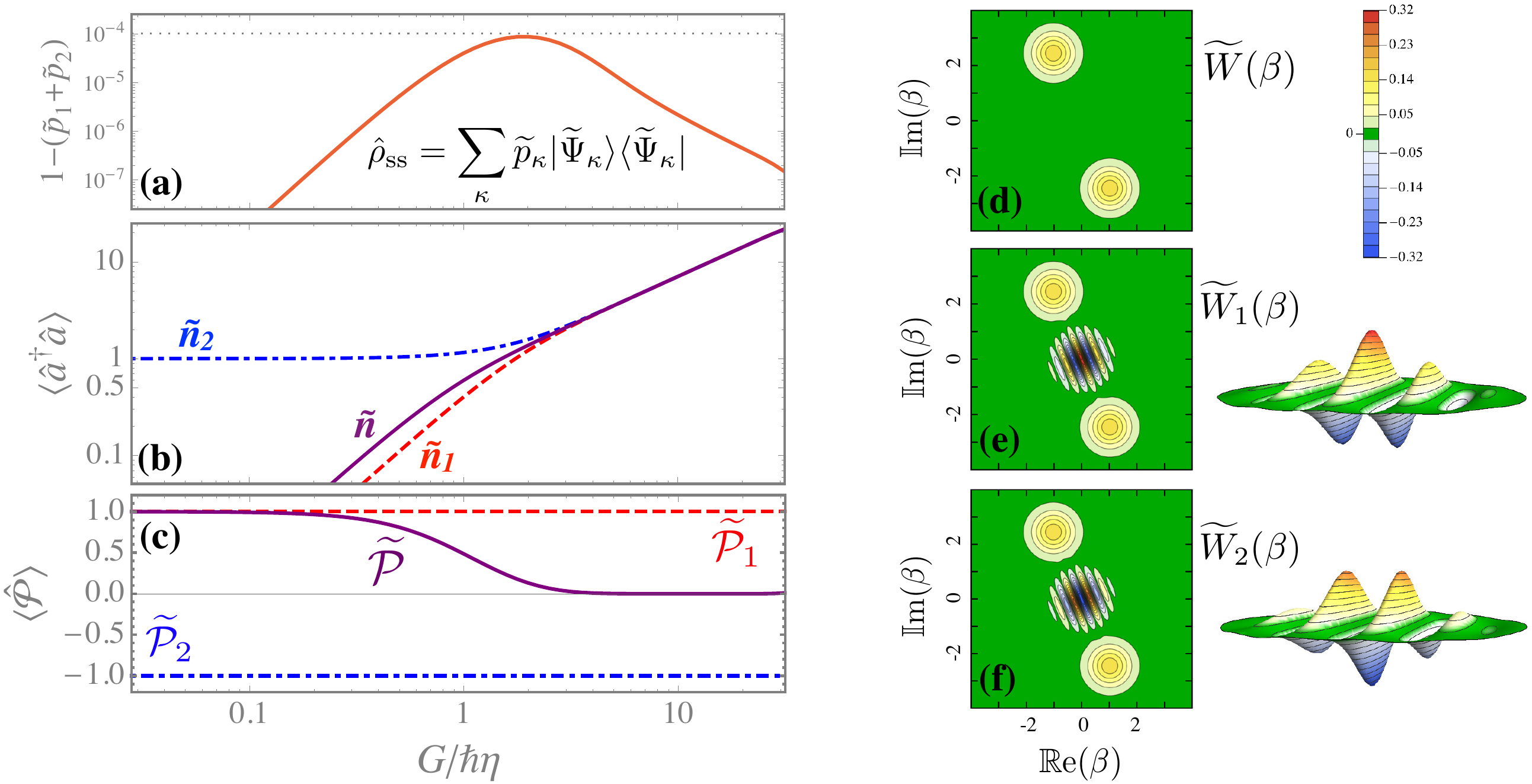}
	\caption{{\bf Exact results for the steady state.} The corresponding density matrix can be expressed as $\hat{\rho}_{\rm ss} = \sum_\kappa \widetilde{p}_\kappa \vert \widetilde{\Psi}_\kappa \rangle \langle \widetilde{\Psi}_\kappa \vert$, where $\widetilde{p}_1$ and $\widetilde{p}_2$ are the probabilities of the two most probable eigenstates.
		Parameters: $\Delta=0$, $U=\hbar \eta$, $\gamma=0.1 \, \eta$.
		Panel~(a): residual probability $1-\widetilde{p}_1-\widetilde{p}_2$ versus the two-photon drive amplitude $G$ normalized to the two-photon loss rate $\eta$, showing that the density matrix is dominated by the first two eigenstates.
		Panel~(b): as a function of $G/\hbar \eta$, mean number of photons $\widetilde{n}$ and its contributions $\widetilde{n}_1$ and $\widetilde{n}_2$.
		Panel~(c): as a function of $G/\hbar \eta$, the mean parity $\widetilde{\mathcal{P}}$ and its contributions $\widetilde{\mathcal{P}}_1$ and $\widetilde{\mathcal{P}}_2$.
		\new{Panel (d): for $G=10 \hbar \eta$, contour plot of the Wigner function $\widetilde{W}(\beta)$ for the density matrix $\hat{\rho}_{\rm ss}$.
			Panel~(e) and~(f): for $G=10 \hbar \eta$, Wigner functions $\widetilde{W}_1(\beta)$ and $\widetilde{W}_2(\beta)$ associated to the two most probable eigenstates.}
		For the latter, we also show a 3D zoom of the central region $\vert \beta \vert\leq 1.6$.
		\new{Note that the fringe pattern changes sign between  $\widetilde{W}_1(\beta)$ and $\widetilde{W}_2(\beta)$.}}
	\label{fig:Analytic}
\end{figure}

\begin{figure}[t!]
	\includegraphics[width=0.93\textwidth]{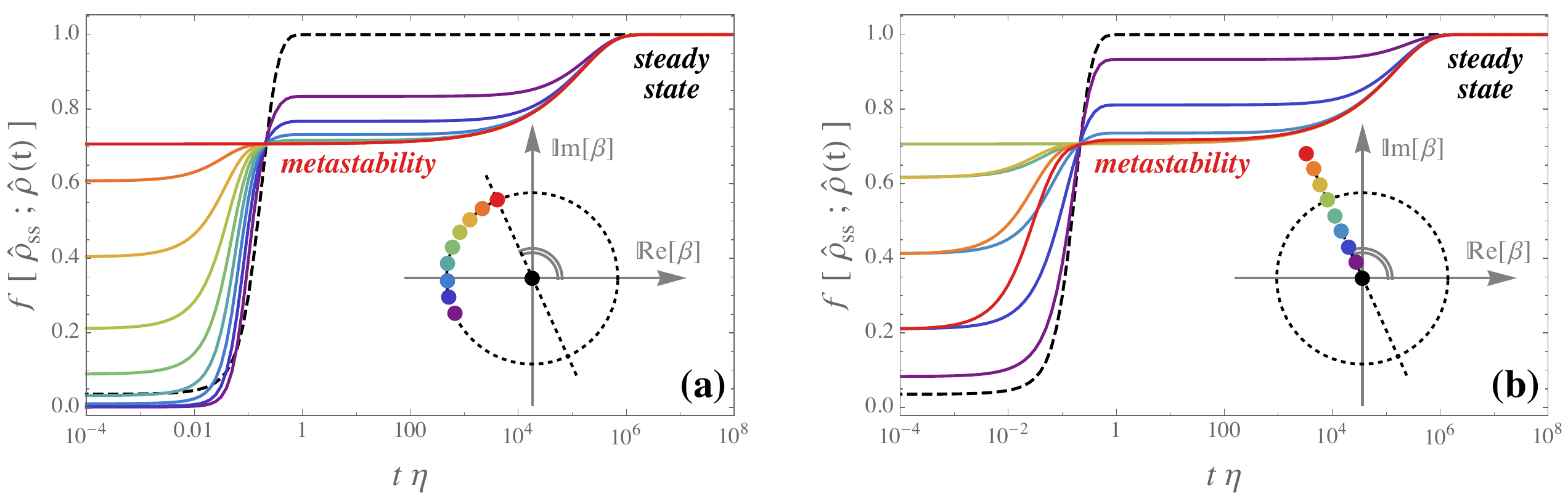}
	\caption{{\bf Metastable versus steady-state regime.} The curves depict the time-dependent fidelity of the density matrix $\hat{\rho}(t)$ with respect to the unique steady-state density matrix $\hat{\rho}_{\rm ss}$ by taking as initial condition a pure coherent state, i.e., $\hat{\rho}(t\!=\!0)=\vert \beta \rangle \langle \beta \vert$.
		The fidelity is defined as $f [\hat{\rho}_{\rm ss};\hat{\rho}(t)] = \Tr {\sqrt{\sqrt{\hat{\rho}_{\rm ss}}\, \hat{\rho}(t)\, \sqrt{\hat{\rho}_{\rm ss}}} }$.
		The values of $\beta$ and the corresponding colours are indicated in the inset.
		\new{Panel~(a): the phase of the initial coherent state is varied (cf. inset).
			Panel~(b): the amplitude is changed (cf. inset).}
		The dashed lines correspond to the vacuum as initial state.
		Parameters: $\Delta=0$, $U=\hbar\eta$, $G=10\hbar\eta$, $\gamma=\eta$.}
	\label{fig:Metastable}
\end{figure}

To further characterise the steady-state, we consider the spectral decomposition of the density matrix \mbox{$\hat{\rho} = \sum_{\kappa} p_{\kappa} \vert \Psi_{\kappa} \rangle \langle \Psi_{\kappa} \vert$}, with $\vert \Psi_{\kappa} \rangle$ the $\kappa^{\rm th}$ eigenstate of $\hat{\rho}$ with eigenvalue $p_{\kappa}$.
The latter corresponds to the probability of finding the system in $\vert \Psi_{\kappa} \rangle$. The eigenstates are sorted in such a way that $p_{\kappa}\ge p_{\kappa +1}$.
For a pure state, $p_1=1$ and all the other probabilities $p_\kappa$ are zero.
\new{We numerically diagonalised the density matrix $\hat{\rho}_{\rm ss}$ in a truncated Fock basis, choosing a cutoff ensuring a precision of $10^{-14}$.
For our calculations, we chose a set of parameters around typical experimental ones~\cite{LeghtasScience15}, i.e. $\Delta\simeq0$, $|U|\simeq\hbar\eta$, $G\gtrsim\hbar\eta$, and $\gamma\lesssim\eta$.
In this regime,} for the steady state~\eqref{eq:Rhonm} only two eigenstates dominate the density matrix.
As shown in Fig.~\ref{fig:Analytic}(a), typically $\widetilde{p}_1+\widetilde{p}_2 \simeq 1$, and
$\hat{\rho}_{\rm ss} \simeq
\widetilde{p}_1 \vert \widetilde{\Psi}_1 \rangle \bra{\widetilde{\Psi}_1}+
\widetilde{p}_2 \vert \widetilde{\Psi}_2 \rangle \bra{\widetilde{\Psi}_2}$.
The aforementioned absence of even-odd coherences implies that $\ket{\widetilde{\Psi}_{1(2)}}$ is composed of only even (odd) Fock states.
Furthermore, we find that $\ket{\widetilde{\Psi}_{1}}\simeq\ket{\mathcal{C}_\alpha^+}$ and $\ket{\widetilde{\Psi}_{2}}\simeq\ket{\mathcal{C}_\alpha^-}$ for an appropriate choice of $\alpha$. 
For the parameters of Fig.~\ref{fig:Analytic}(d),  $\braket{\widetilde{\Psi}_{1(2)}}{\mathcal{C}_\alpha^{+(-)}}\simeq (1-8\times10^{-6})$ for $\alpha \approx 2.7\, e^{2.0\,i}$.
\new{We have varied $\Delta/\hbar\eta$ between -0.2 and 0.2, $G/\hbar\eta$ between 0 and 15, $\gamma/\eta$ between 0 and 5, $U/\hbar\eta$ between 1 and 10, always finding that $1-\tilde{p}_1-\tilde{p}_2 < 10^{-2}$.
	Moreover, in these ranges, we verified that there exists a value of $\alpha$ such that $\braket{\widetilde{\Psi}_{1(2)}}{\mathcal{C}_\alpha^{+(-)}}>0.98$.
	Hence, we can conclude that for a broad range of parameters the eigenstates of $\hat{\rho}_{\rm ss}$ are two cat-like states of opposite parity.}

Using the linearity of the trace, for any operator $\hat{O}$ one can write
$\widetilde{O}=\Tr{\hat{\rho}_{\rm ss}\, \hat{O} }\simeq \widetilde{p}_1 \widetilde{O}_1+\widetilde{p}_2 \widetilde{O}_2$,
where $\widetilde{O}_\kappa=\mean{\widetilde{\Psi}_\kappa}{\hat{O}}{\widetilde{\Psi}_\kappa}$.
In Fig.~\ref{fig:Analytic}(b) we plot, as a function of the pump amplitude $G$, the steady-state mean density $\widetilde{n}$, together with its contributions $\widetilde{n}_{1,2}$.
\new{For weak pumping one has $\widetilde{n}_{1}\simeq 0$ and $\widetilde{n}_{2}\simeq 1$, in agreement with what one would obtain for the even and the odd cat by taking the limit $\alpha \to 0$ of Eq.~\eqref{eq:catdef}.}
These two contributions become equal in the limit of a very large number of photons.
As shown in Fig.~\ref{fig:Analytic}(c), the two contributions to the mean parity $\widetilde{\mathcal{P}}_{1,2}$ always stay clearly different, being $\ket{\mathcal{C}_\alpha^\pm}$ orthogonal eigenstates of $\hat{\mathcal{P}}$ with eigenvalues $\pm1$.
A valuable tool to visualise the nonclassicality of a state is the Wigner function $W(\beta)=\frac{2}{\pi}\Tr{\hat{\rho}\, \hat{D}_\beta \hat{\mathcal{P}} \hat{D}_\beta^\dagger}$, defined in terms of the displacement operator~\cite{WignerPR32} $\hat{D}_\beta=\exp(\beta \hat{a}^\dagger - \beta^* \hat{a})$.
Indeed, negative values of $W(\beta)$ indicate strong non-classicality~\cite{HarocheBOOK}.
The Wigner function corresponding to the density matrix~\eqref{eq:Rhonm} is always positive [cf. Eq.~\eqref{eq:wigner} in Methods], while the separate contributions $\widetilde{W}_1(\beta)$ and $\widetilde{W}_2(\beta)$ exhibit an interference pattern with negative regions, typical of cat states [cf. Fig.~\ref{fig:Analytic}(d-f)].

We emphasize that for  finite $\gamma$ the considered system has always a unique steady state.
However, the temporal relaxation towards the steady state depends dramatically on the initial \new{condition}.
This is revealed by the time-dependent fidelity with respect to the steady state, presented in Fig.~\ref{fig:Metastable}, obtained by numerical integration of the master equation. 
In particular, initialising the system in one of the coherent states $\vert \pm \alpha \rangle$ composing the steady-state cats, it persists nearby for a time several orders of magnitude longer than $1/\gamma$ and $1/\eta$. 
Hence, the ``multiple stable steady states'' in~\cite{LeghtasScience15} are actually metastable. 

\subsection*{Quantum trajectories}

\begin{figure}[t!]
	\includegraphics[width=0.71 \textwidth]{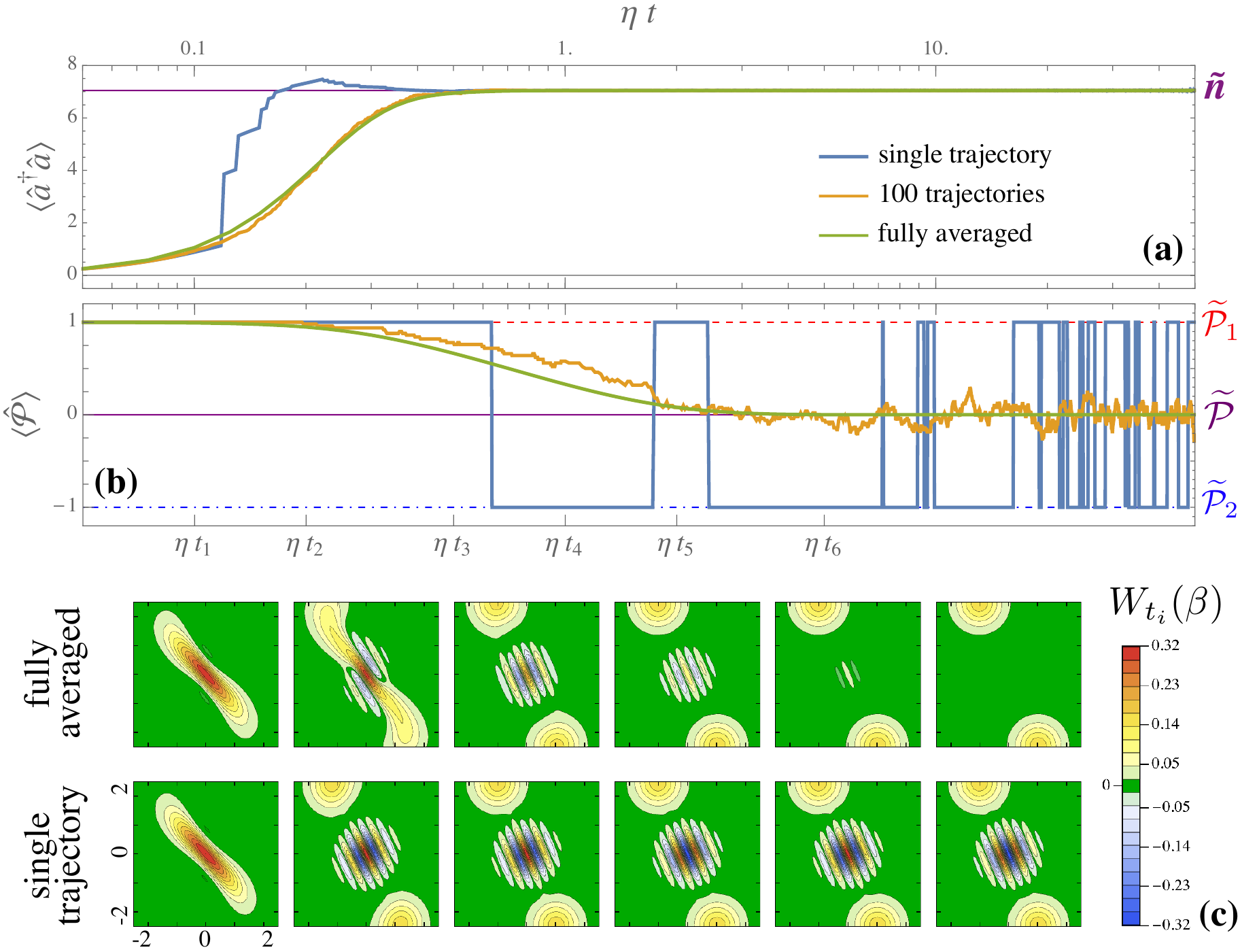}
	\caption{{\bf Dynamics of averaged quantities versus single quantum trajectories.} Panel~(a): dynamics (time is in logarithmic scale) of  the photon population for a single quantum trajectory (blue line), an average of $100$ trajectories (orange line) and the fully averaged (green line) density matrix.
		Panel~(b): same as~(a) but for the expectation value of the photon parity operator.
		Panel~(c): snapshots of the Wigner functions at different times.
		The system parameters are $\Delta=0$, $U=\hbar\eta$, $G=10\hbar\eta$, and $\gamma=0.1\eta$.
		\new{We stress that, in the single-trajectory Wigner functions, the fringe pattern changes sign after a parity jump.}}
	\label{fig:Trajectories}
\end{figure}

We now examine the quantum trajectories of the system, which give an insight on the pure states that the system explores during its dynamics~\cite{CarmichaelPRL93,MolmerJOSAB93,PlenioRMP98}.
In principle, keeping track of all the photons escaping the cavity allows to follow the system wave function (cf. Methods)~\cite{CarmichaelPRL93,GardinerBOOK}.
A \new{photon-counting} trajectory is presented in Fig.~\ref{fig:Trajectories}, where in panels~(a,b) we follow, respectively, the time evolution of the photon number $\mv{\hat{n}}$ and of the parity $\mv{\hat{\mathcal{P}}}$, starting from the vacuum state as initial condition.
On a single trajectory, two-photon processes initially dominate, driving the system towards $\ket{\mathcal{C}_\alpha^+}$.
Indeed, $\mv{\hat{n}}\simeq \widetilde{n}_1$ and $\mv{\hat{\mathcal{P}}}=\widetilde{\mathcal{P}}_1=1$.
Two-photon losses do not affect a state parity, indeed $\hat{a}^2\,\ket{\mathcal{C}_\alpha^\pm}=\alpha^2\,\ket{\mathcal{C}_\alpha^\pm}$.
This is why the system persists nearby the even cat until a one-photon loss occurs.
At this point, the state abruptly jumps to the odd manifold~\cite{KrippnerPRA94}, since $\hat{a}\,\ket{\mathcal{C}_\alpha^\pm}\propto\ket{\mathcal{C}_\alpha^\mp}$.
After the jump, two-photon processes stabilise $\ket{\mathcal{C}_\alpha^-}$, so that $\mv{\hat{n}}\simeq \widetilde{n}_2$ and $\mv{\hat{\mathcal{P}}}=\widetilde{\mathcal{P}}_2=-1$.
When another one-photon jump takes place, the system is brought back to the even manifold, and so on.
Hence, if the quantum trajectory is monitored via photon counting~\cite{GardinerBOOK}, the system can only be found nearby $\ket{\mathcal{C}_\alpha^+}$ or $\ket{\mathcal{C}_\alpha^-}$.
The probability of being in each cat is given by the corresponding eigenvalue of $\hat{\rho}_{\rm ss}$, namely $\widetilde{p}_1$ and $\widetilde{p}_2$.
Since $\widetilde{n}_{1,2}\approx\widetilde{n}$, it is impossible to discern the cats' jumps by tracking the photon density.
A parity measurement, instead, would be suitable \cite{SunNature2014}.
In Fig.~\ref{fig:Trajectories}(a) and~(b) we also \new{present} the average over 100 trajectories, which, as expected, converges to the master equation solution \new{(also shown)}.
The latter corresponds to the full average over an infinite number of realizations~\cite{CarmichaelBOOK}.
The fully-averaged and single-trajectory evolutions of the Wigner function are shown in Fig.~\ref{fig:Trajectories}(c).
In the averaged one, an even-cat transient appears, but negativities are eventually washed out for $\eta t,\gamma t\gg1$~\cite{GillesPRA94,KrippnerPRA94,LeghtasScience15}.
By following a single quantum trajectory, instead, we see that $W_t(\beta)$ quickly tends to the one of $\ket{\mathcal{C}_\alpha^+}$.
Then, it abruptly switches to that of $\ket{\mathcal{C}_\alpha^-}$, then back at each one-photon jump. 

\begin{figure}[t!]
	\includegraphics[width=0.70 \textwidth]{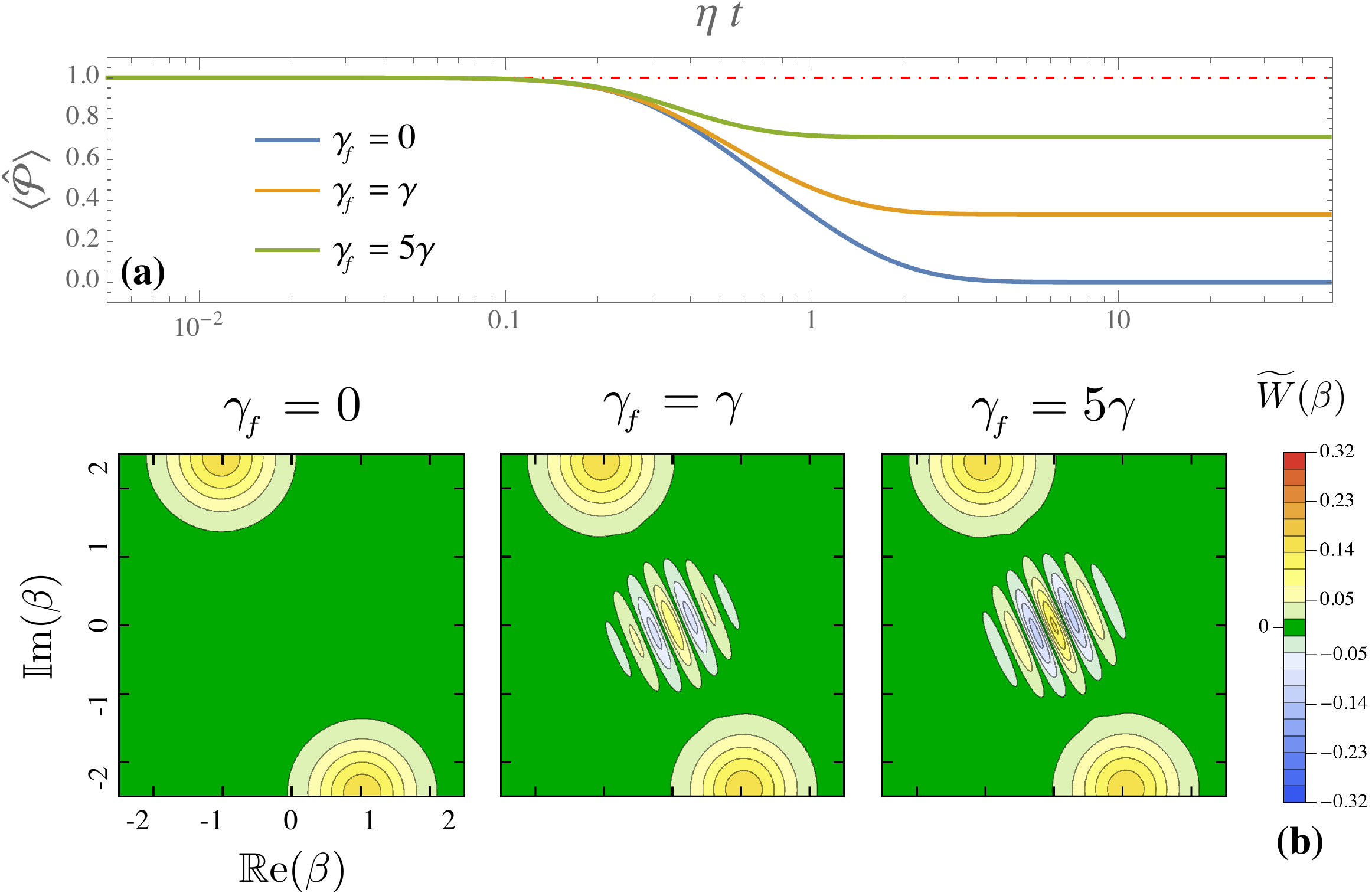}
	\caption{
		{\bf Effects of the considered feedback.} These results are obtained by \new{numerical integration of the Lindblad master equation,} taking the vacuum as initial condition and for same system parameters as in Fig.~\ref{fig:Trajectories}.
		Panel~(a): time evolution of the parity mean-value in presence of the feedback described by Eq.~\eqref{eq:demon} for different values of $\gf$ (cf. legend).
		Panel~(b): steady-state Wigner function ($t\to + \infty$) exhibiting negativities for $\gf>0$.}
	\label{fig:Feedback}
\end{figure}

\new{The even-to-odd jumps in a photon-counting quantum trajectory deserve a more detailed discussion.
	Each trajectory corresponds to the behaviour of our quantum system on  a single-shot experiment~\cite{CarmichaelBOOK,CarmichaelPRL93}.
	Indeed, to simulate a quantum trajectory it is necessary to model how an observer measures the environment to probe the system (in the presented case, by a photon-counting measure).
	The same Lindblad equation can be described via different measurement protocols, resulting in different single trajectories. 
    Their average result reproduces the master equation solution~\cite{HarocheBOOK,MolmerJOSAB93,PlenioRMP98}.
	Our steady state, given by $\hat{\rho}_{\rm ss}$ in Eq.~\eqref{eq:Rhonm}, is typically a mixture of an even and an odd cat state.
	To unveil such a mixture at the single-trajectory level, a photon-counting measurement is suitable.
	But this does not mean that \emph{any} physical system described by the same Lindblad equation would \emph{always} be in a parity-defined state.
	We emphasise that this behaviour is not exclusively caused by the chosen measurement process: other systems under the same photon-counting trajectory do not show even-to-odd jumps.
	For example, if one considers a one-photon pump and no two-photon driving, at any given time the trajectories do not show a defined parity (the state of the system in not an eigenstate of $\hat{\mathcal{P}}$).}

\subsection*{A feedback protocol}
\new{The results presented above} suggest that, in order to have a cat-like steady state (e.g. keep interference fringes in the fully averaged Wigner function), one may try to unbalance the even and odd contributions to $\hat{\rho}_{\rm ss}$.
This effect can be envisioned through a parity-triggered feedback mechanism \cite{VitaliJMO04,ZippilliPRA03,SayrinNature11,ZhouPRL12} \new{opening} a one-photon loss channel.
\new{In practice, this can be implemented via non-destructive parity measurements~\cite{CampagnePRX13, SunNature2014}, whose rate must be larger than any other rate in the system.
	Note that, in general, a parity measurement projects the system into the even- or odd-parity manifold, affecting the dynamics by destroying the even-odd coherences.
	In the present case, however, those coherences are proven to be always zero in the steady state by the analytic solution~\eqref{eq:Rhonm}.
	Thus, a high-rate and non-destructive parity measure does not alter significantly the system dynamics and allows to continuously monitor $\mv{\hat{\mathcal{P}}}$.}
When the undesired value is measured, an auxiliary qubit is put into resonance with the cavity, inducing the absorption of a photon. 
After the desired parity is restored, the qubit is brought out of resonance, closing the additional dissipation channel.
\new{Such a qubit acts as a non-Markovian bath for the system, and in principle its effects can not be simply assimilated to those of a Markovian environment.
However, if one assumes that the excited-state lifetime of the qubit is shorter than the inverse of the qubit-cavity coupling rate, one can safely treat it as an additional Markovian dissipator~\cite{GoetschPRA96,ZippilliPRA03}.
In other words, the coupled qubit must be engineered to easily lose the photon to the environment, which seems a reasonable task for the present experimental techniques~ \cite{BrunePRA92,SayrinNature11,ZhouPRL12,EverittFICT14}.
Under these assumptions, the proposed feedback protocol} can be effectively described by the \new{additional} jump operator $\af=\hat{a}\,\frac{1}{2}(1-\hat{\mathcal{P}})$ and the corresponding dissipator
\begin{equation}\label{eq:demon}
	\Df \hat{\rho}= \frac{\gf}{2} \left(2 \af \hat{\rho} \af^{\dagger} -\af^{\dagger}\af \hat{\rho} - \hat{\rho} \af^{\dagger} \af \right).
\end{equation}
Qualitatively, $\Df$ leaves the even cat undisturbed, while it enhances the dissipation for the odd one.

In Fig.~\ref{fig:Feedback}(a) we show the time evolution of $\langle\hat{\mathcal{P}}\rangle$ for three different values of $\gf$.
These results have been \new{obtained via numerical integration of the Lindblad master equation with a total dissipator $\mathcal{D}=\mathcal{D}_1+\mathcal{D}_2+\mathcal{D}_f$}.
At the steady state, as $\gf$ increases so does $\widetilde{\mathcal{P}}$, indicating that the positive cat has a larger weight in $\hat{\rho}_{\rm ss}$.
In Fig.~\ref{fig:Feedback}(b) we show the corresponding steady-state Wigner functions  $\widetilde{W}(\beta)$.
For finite $\gf$, negative fringes appear in the Wigner function. They are more pronounced as $\gf$ is increased, revealing a highly nonclassical state.
In the limit $\gf\gg\gamma$,  $\hat{\rho}_{\rm ss}\simeq \ket{\mathcal{C}_\alpha^+}\bra{\mathcal{C}_\alpha^+}$.
By using, instead, the jump operator $\af=\hat{a}\,\frac{1}{2}(1+\hat{\mathcal{P}})$, one can similarly stabilize the odd cat state.
\new{Note that the Wigner-function negativities in Fig.~\ref{fig:Feedback} are those of the full steady-state density matrix. 
Hence, the quantum state of the system is on average nonclassical.}

\section*{Discussion}

In conclusion, we presented the exact steady-state solution for a general photonic resonator subject to one-photon losses and two-photon drive and dissipation.
Remarkably, the unique steady state appears to be a mixture of two orthogonal cat states of opposite parity.
We have also shown that the transient dynamics to the unique steady state can depend dramatically on the initial condition, revealing the existence of metastable states.
Furthermore, by monitoring the quantum trajectory of the system via photon counting, we found that it explores the two cat states composing the steady-state statistical mixture.
On this ground, we proposed to engineer a parity-dependent dissipation which allows to stabilize a cat-like steady state.

The general nature and richness of the results predicted here paves the way to a wide variety of experimental and theoretical investigations.
As a future perspective, a challenging but intriguing problem is the study of other photonic cat-like states in the transient and steady-state regime for arrays of coupled resonators.
The generation and stabilization of orthogonal cat-like states is of great interest for quantum computation, since they can be used as qubits logic states~\cite{GilchristJOB04,OurjoumtsevScience06,MirrahimiNJP14}.
Besides the implications for quantum information, our results are also relevant for the study of exotic phases based on the manybody physics of light~\cite{CarusottoRMP13}.

\section*{Methods}

\subsection*{Complex P-representation}
The complex $P$-representation of the density matrix $\hat{\rho}$ reads~\cite{DrummondJPA80}
\begin{equation}\label{eq:generalP}
	\hat{\rho}=\int_{\mathcal{C}} \int_{\mathcal{C'}}
	\frac{\ket{\alpha}\bra{\beta^*}}
	{\braket{\beta^*}{\alpha}}\,
	P(\alpha, \beta)\,d\alpha\,d\beta,
\end{equation}
where the complex amplitudes $\alpha$ and $\beta$ define the corresponding (nonorthogonal) coherent states.
The two independent and closed integration paths $\mathcal{C}$ and $\mathcal{C'}$ must encircle all the singularities of $P(\alpha,\beta)$ in the complex plane.
A Lindblad master equation for the density matrix translates into a Fokker-Plank-like differential equation for the function $P(\alpha, \beta)$~\cite{DrummondOA81}.
In the case defined by Eqs.~\eqref{eq:hamiltonian2phot_a}-\eqref{eq:hamiltonian2phot_b}-\eqref{eq:hamiltonian2phot_c}, one has 
\begin{equation}
	\begin{split}
		\frac{\partial P(\alpha, \beta)}{\partial t} =
		\left[\frac{\partial}{\partial \alpha} \left(\frac{\hbar \gamma}{2} \alpha \new{- i \Delta \,\alpha} + i \,  \mathcal{U} \alpha^2 \beta +  i G \beta\right)+ 
		\frac{\partial}{\partial \beta} \left( \frac{\hbar \gamma}{2} \beta \new{+ i \Delta\, \beta} - i  \, \mathcal{U}^* \alpha \beta^2 -  i G^* \alpha \right) \right. \\
		\left.-\frac{i}{2}  \frac{\partial^2}{\partial \alpha^2}  \left(\mathcal{U}\,\alpha^2 + G\right)+
		\frac{i}{2}  \frac{\partial^2}{\partial \beta^2} \left(\mathcal{U}^*\,\beta^2 + G^*\right) \right]
		P(\alpha, \beta),
	\end{split}
\end{equation}
where $\mathcal{U}=U - i \hbar \eta$.
The corresponding steady-state equation, defined by $\partial_t \widetilde{P}(\alpha, \beta)=0$, is satisfied by
\begin{equation}
	\widetilde{P}(\alpha,\beta)
	\propto
	\frac{e^{2\,\alpha\,\beta}}
	{(\alpha^2+g)^{1+c}(\beta^2+g^*)^{1+c^*}},
\end{equation}
\new{where $c$ and $g$ were introduced in the main text.}
Taylor expanding the exponential and projecting on the number states $\bra{n}$ and $\ket{m}$, we obtain a formal expression for the density matrix elements of $\hat{\rho}_{\rm ss}$:
\begin{equation}\label{eq:rhocoeff}
	\bra{n}\hat{\rho}_{\rm ss}\ket{m}=\frac{1}{\mathcal{N}} \sum_{\ell=0}^{\infty} \frac{1}{\ell! \sqrt{n! m!}} \,
	\int_{\mathcal{C}} d\alpha \frac{\alpha^{n+\ell}}{\left( \alpha^2 + g \right)^{1+c}} \int_{\mathcal{C}} d{\beta} \frac{\beta^{m+\ell}}{\left( \beta^2 + g^* \right)^{1+c^*}}.
\end{equation}
The appropriate choice of the integration paths $\mathcal{C}$ is a central issue.
A suitable one is the Pochhammer contour~\cite{DrummondOA81}, which leads to Eq.~\eqref{eq:Rhonm}.
Similarly, it is also possible to calculate the general steady-state expectation value of the correlation functions
\begin{equation} \label{eq:momenta}
	\langle\hat{a}^{\dagger\,n}\hat{a}^m\rangle=
	\frac{1}{\mathcal{N}}\,
	\sum_{\ell=0}^{\infty} \frac{2^\ell}{\ell!}\,
	\mathcal{F}(g,c,\ell+m)\,\mathcal{F}^*(g,c,\ell+n),
\end{equation}
and the steady-state Wigner function (without feedback)
\begin{equation}\label{eq:wigner}
	\widetilde{W}(\beta)=\frac{2}{\pi\mathcal{N}}\,e^ {-2 |\beta|^2}\,
	\left|\sum_{\ell=0}^{\infty} \frac{(2\beta^*)^{\ell}}{\ell!} \mathcal{F}(g,c,\ell)\right|^2.
\end{equation}

\subsection*{Determination of quantum trajectories}
The Lindblad master equation defined by Eqs.~\eqref{eq:hamiltonian2phot_a}-\eqref{eq:hamiltonian2phot_b}-\eqref{eq:hamiltonian2phot_c} describes the evolution of the density matrix if we do not collect any information about the system or the environment.
Hence, the analytic and numerical solution of a Linbdblad master equation predicts the average outcomes of an experimental realisation.
Detecting the photons escaping the system would provide more information about the time evolution of the system itself~\cite{MolmerJOSAB93,PlenioRMP98,GardinerBOOK}.
That is, the evolution of the wave function of an open quantum system can be followed gathering information on its exchanges with the environment.
The operator $\hat{\mathcal U}(t)=\exp(-i\,t\,\hat{H}_{\rm eff}/\hbar)$ describes the time evolution of the system state between two detections of a photon emission.
\new{For our system, the non-hermitian effective Hamiltonian reads:}
\begin{equation}
	\hat{H}_{\rm eff}=\hat{H}
	-i\hbar\frac{\gamma}{2}\hat{a}^\dagger\hat{a}
	-i\hbar\frac{\eta}{2}\hat{a}^\dagger\hat{a}^\dagger\hat{a}\hat{a}.
\end{equation}
At $t=t_i$, the emission of $\nu_i$ photons is detected ($\nu_i=1,2$) and the state abruptly changes according to $\ket{\psi(t_i+\delta t)}\propto\hat{a}^{\nu_i}\ket{\psi(t_i)}$.
This is a quantum jump, which can be simulated stochastically: in the interval $]t,t+\delta t]$ the probability of one-photon emission is $\propto\delta t\,\gamma\,\mean{\psi(t)}{\hat{a}^\dagger\hat{a}}{\psi(t)}$, while that of two-photon emission is  $\propto\delta t\,\eta\,\mean{\psi(t)}{\hat{a}^{\dagger 2}\hat{a}^2}{\psi(t)}$.
After a jump, the time evolution continues under the action of $\hat{\mathcal U}(t-t_i)$ until the next photon emission.
On this ground, a single trajectory is simulated by randomly determining if, at each time step, the state jumps or evolves under the action of $\hat{\mathcal U}$.
We stress that for perfect detection (all the emitted photons are gathered) and a pure initial condition, the state $\ket{\psi(t)}$ stays pure at any time.
\new{For the same Lindblad equation, other quantum trajectories than the photon-counting ones can be modelled and simulated~\cite{HarocheBOOK,CarmichaelBOOK,GardinerBOOK}.
	However, the average over an infinite number of trajectories will always give the solution of the Lindblad master equation.}

{\footnotesize

}

\section*{Acknowledgements}
We thank B.~Huard, Z.~Leghtas, and G.~Rembado for discussions. We  acknowledge  support  from  ERC  (via  the Consolidator  Grant ``CORPHO'' No.  616233) and from ANR (Project QUANDYDE No. ANR-11-BS10-0001).

\section*{Author contributions}
F.M., N.B., W.C. obtained the exact solution for the steady-state density matrix. N.B. and F.M. performed the time-dependent solutions of the master equation, while F.M., J.L. studied the individual quantum trajectories. N.B. composed the figures. All authors contributed to the critical analysis of the results and writing of the manuscript. C.C. proposed and supervised the project. 

\section*{Competing financial interests}
The authors declare no competing financial interests.

\end{document}